\documentclass[aps,prl,twocolumn,showpacs,superscriptaddress,amsmath,amssymb]{revtex4}
\usepackage{epsfig}
\usepackage{graphicx}
\usepackage{citesort}

\begin{document}

\title{{\rm\small\hfill (Phys. Rev. Lett., in press)}\\
Dissociation of O$_2$ at Al(111): The Role of Spin Selection Rules}

\author{J\"org Behler}
\affiliation{Fritz-Haber-Institut der Max-Planck-Gesellschaft, Faradayweg 4-6, D-14195 Berlin, Germany}
\author{Bernard Delley}
\affiliation{Paul-Scherrer-Institut, HGA/123, CH-5232 Villigen PSI, Switzerland}
\author{S\"onke Lorenz}
\affiliation{Fritz-Haber-Institut der Max-Planck-Gesellschaft, Faradayweg 4-6, D-14195 Berlin, Germany}
\author{Karsten Reuter}
\affiliation{Fritz-Haber-Institut der Max-Planck-Gesellschaft, Faradayweg 4-6, D-14195 Berlin, Germany}
\author{Matthias Scheffler}
\affiliation{Fritz-Haber-Institut der Max-Planck-Gesellschaft, Faradayweg 4-6, D-14195 Berlin, Germany}

\received{28 October 2004}

\begin{abstract}
A most basic and puzzling enigma in surface science is the description of the dissociative adsorption of O$_2$ at the (111) surface of Al. Already for the sticking curve alone, the disagreement between experiment and results of
state-of-the-art first-principles calculations can hardly be more dramatic. In this paper we show that this is caused by hitherto unaccounted spin selection rules, which give rise to a highly non-adiabatic behavior in the O$_2$/Al(111) interaction. We also discuss problems caused by the insufficient accuracy of present-day exchange-correlation functionals.
\end{abstract}


%% 82.20.Kh Potential energy surfaces for chemical reactions
%% 82.20.Gk Electronically non-adiabatic reactions
%% 68.35.Ja Surface and interface dynamics and vibrations 

\maketitle

Oxygen-metal interactions are responsible for everyday phenomena 
like corrosion, and form the atomic-scale basis behind numerous 
technological applications like oxidation catalysis. It is therefore most 
discomforting that despite several decades of research in surface science, 
the initial step in the oxygen-metal interaction, namely the dissociation 
process of O$_2$ molecules over metal surfaces, is not yet understood. This 
is in particular so for what is often called the most simple metal surface, 
namely Al(111): a close-packed surface of a nearly-free electron metal. 
For the initial interaction of O$_2$ with Al(111) experiments have 
consistently shown \cite{osterlund97,brune92} that the initial dissociative 
sticking probability for thermal O$_2$ is very low (about 2\%). 
Density-functional theory (DFT) calculations, on the other hand, found that 
dissociation is not hindered by energy barriers \cite{honkala00}, which 
implies that the initial sticking coefficient should be very high (about 
100\%). Another intriguing aspect of the O$_2$/Al(111) system is that
at very low coverages the distribution of adsorbed oxygen atoms is random, 
even when adsorption is performed at temperatures at which thermal
diffusion can not play a significant role~\cite{brune92}. Thus, it is 
impossible to trace back which two adatoms stem from the same molecule. 
Initially this led to the suggestion that the adsorption energy is used to 
trigger the diffusion of ``hot adatoms''~\cite{brune92}. More recently, a 
different explanation has been suggested (``abstraction''), where only one 
O-atom is adsorbed and the other one is repelled back into the 
vacuum~\cite{komrowski01}. Again, theoretical work, so far, does not give 
a clue why this may be so. Thus, one may ask, what we can trust in surface 
science when understanding of such a most basic and simple system for 
molecule-surface interactions is so clearly lacking.

\begin{figure}[t]
\scalebox{0.23}{\includegraphics{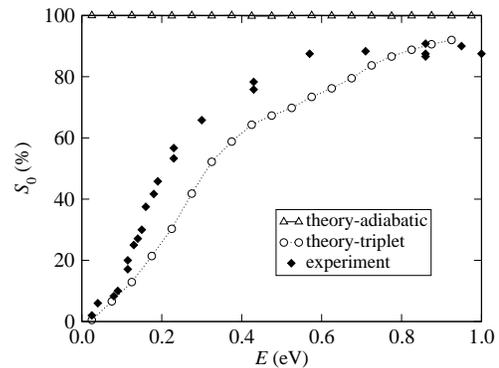}} 
\caption{\label{sticking} Initial sticking curve of O$_2$ at Al(111), 
based on the adiabatic (empty triangles) and the spin-triplet (empty circles) 
potential-energy surfaces using the RPBE functional. The experimental 
data (solid diamonds) are from ref. \cite{osterlund97}.}
\end{figure}

Figure \ref{sticking} summarizes the experimental data for the initial 
sticking coefficient as function of the kinetic energy of
incoming O$_2$ molecules for a molecular beam at normal incidence (full diamonds) \cite{osterlund97}, as well as the result, of what has hitherto 
been the standard theoretical treatment (labeled as ``theory-adiabatic''). 
Also shown is the result of the approach taken in the present paper (labeled 
as ``theory-triplet''), which will be detailed below. Obviously, there is 
hardly any similarity between the ``theory-adiabatic'' curve and the 
experimental result. Though we called this the ``standard theoretical 
treatment'', we note that already the calculations behind the 
``theory-adiabatic'' curve (and also behind the ``theory-triplet'' 
curve) are much more elaborate and advanced than typical approaches 
to obtain the initial sticking coefficient: 
All theoretical results presented in this paper were obtained from extensive
all-electron DFT calculations using the DMol$^3$ code~\cite{delley90}. This 
provided the six-dimensional potential-energy surface (PES) for the
O$_2$/Al(111) system at more than 1500 geometries of the two oxygen atoms, 
keeping the substrate frozen. These PES data points were then interpolated
by a neural-network~\cite{lorenz04,tobepublished}, enabling us to perform
molecular dynamics (MD) calculations for about 100,000 trajectories, including
all possible initial molecular orientations. Thus,
this approach~\cite{gross97} grants a controlled and good statistics, in 
contrast to ``on-the-fly {\em ab initio} MD'', which gives (for a frozen 
substrate) the same trajectories, but where due to the high CPU cost at best 
only $\sim 50$ trajectories could be performed even on todays biggest 
computers.

Still, ``on-the-fly {\em ab initio} MD'' has the advantage that it can also
be used beyond the frozen substrate approximation. To check on the
validity of our treatment, we therefore performed 24 {\em ab initio} MD 
runs, where the full dynamics of the Al surface atoms was taken into account. 
These studies show that the adsorption energy is efficiently transferred to 
strong surface vibrations, and that the oxygen adatoms do not move far. Thus,
the ``hot adatom'' concept is not supported. In all studied trajectories 
the Al(111) surface got only affected, when the O$_2$ was quite close to 
the surface, i.e. when O-Al bonds were being formed and the O-O bond notably 
weakened (at molecule-surface distances below $\approx 2.5$~\AA). Before
this point, the O$_2$ trajectories were not changed by the substrate
vibrations, and in particular all incoming O$_2$ molecules are found to 
dissociate, fully confirming the adiabatic result shown in Fig. 
\ref{sticking}. We also performed a systematic comparison using different 
exchange-correlation (xc) functionals, including the PBE~\cite{perdew96} 
and RPBE~\cite{hammer99}. The resulting PESs look different in some details, 
however, the resulting sticking curve is always essentially the same as the 
``theory-adiabatic'' curve in Fig. \ref{sticking}. Hence, neither the 
approximate xc treatment, nor the frozen substrate approximation can
account for the dramatic disagreement between the theoretical and experimental 
results. We therefore conclude that the origin must be more fundamental, 
namely in the assumed adiabatic description, restricting the impinging 
molecule to the electronic ground state of the combined O$_2$/Al system at
each point of the O$_2$ trajectory. Based on less rigorous studies, this had 
been suggested previously~\cite{hellman03, binetti03}.

Inspecting the six-dimensional adiabatic PES reveals immediately an 
obvious flaw of the adiabatic description, independent of the employed
xc functional: Even at largest distances the electron chemical potentials 
of the O$_2$ molecule and the Al(111) surface align, which is achieved by 
some electron transfer towards the O$_2$ molecule. Obviously, in reality 
charge transfer will occur only when the two systems are getting close for 
a sufficiently long period of time. Recently, Hellman {\em et 
al.}~\cite{hellman03} considered the influence of charge transfer by 
employing an approach, where they replaced the Al(111) surface by jellium 
and treated the kinetic-energy operator in the Thomas-Fermi-Weizs\"acker
approximation. Then, two one-dimensional diabatic PESs were constructed, 
one where the O$_2$ molecule was kept neutral and one where a full electron
was transferred~\cite{hellman03}. This description could indeed account 
for the qualitative shape of the experimental sticking curve, as could 
Binetti {\em et al.} \cite{binetti03}, following a comparable approach,
but considering four different diabatic model PESs. Both treatments 
point therefore at the possible importance of non-adiabatic effects, but 
due to the arbitrary and severe approximations, doubts remain about their
conclusiveness.

Our works starts from recognizing that chemical interactions are ruled by 
various selection rules, and for the present situation spin-conservation 
~\cite{wigner27} is expected to be relevant. In gas-phase chemistry it is 
well known that O$_2$, when in its triplet ground state, is rather inert 
when the other reactant and the product are spin singlets. Interestingly, 
this role of the O$_2$ spin has not attracted much attention in the 
O$_2$/Al(111) interaction, although it was e.g. studied for the adsorption
of oxygen on Si(100)~\cite{kato98}. The appropriate theoretical modeling 
should then constrain the spin to the O$_2$ Hilbert subspace, preventing 
charge transfer, as well as spin quenching before the systems interact. 
Such a spin-constrained DFT approach has neither been formulated nor 
evaluated for molecule-surface scattering so far. We will show that 
it not only gives a good description of the sticking coefficient (cf. Fig. 
\ref{sticking}, empty circles), but may also explain the enigmatic 
abstraction mechanism.

Let us briefly describe the theoretical method enabling us to study the 
dynamics of an O$_2$ molecule that remains in its spin-triplet configuration.
Only very close to the surface transitions to other configurations of the 
O$_2$/Al(111) system may set in. In order to calculate the spin-triplet 
PES we follow the work of Dederichs {\em et al.} \cite{dederichs84}, for 
which one must first define the Hilbert subspace of the O$_2$ molecule. 
As the DMol$^3$ code employs an atom-centered basis set, we use for this
all basis functions that are also needed to provide a good description of 
the free O$_2$ molecule. Then, for any position of the O$_2$ molecule, we
request that the total electron spin in this Hilbert subspace is one. In 
practice this approach involves the self-consistent filling of the four 
partial densities of states of the spin-up and spin-down, O$_2$ and Al(111)
sub-systems. This is formulated in terms of an auxiliary field in order 
to properly include the effect of the spin-constraint on the total energy 
\cite{tobepublished}.

\begin{figure*}[t]
{\includegraphics[width=16.5cm]{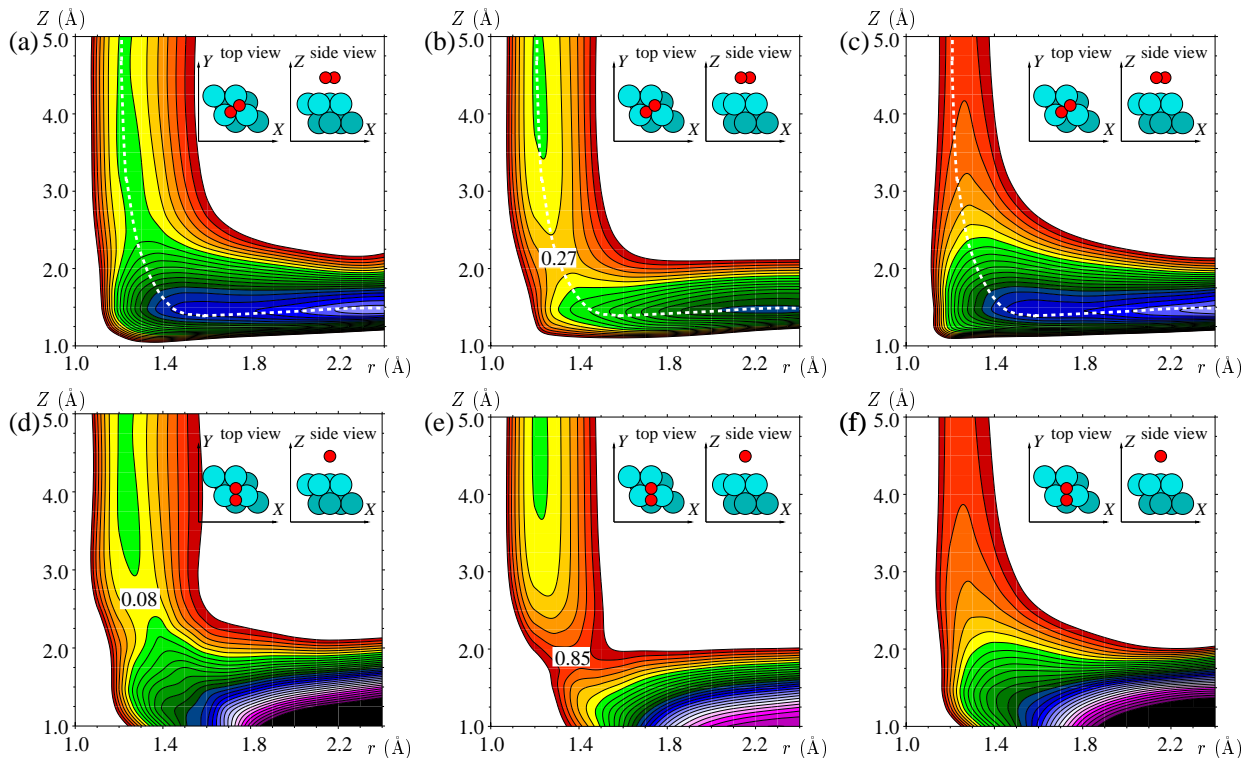}} 
\caption{\label{pes} Two-dimensional (elbow) cuts through the six-dimensional PESs calculated for three different situations, always using DFT-RPBE (see text): adiabatic ($a,d$), triplet ($b,e$) and singlet PES ($c,f$). The 
energies are shown as a function of the O$_2$ bond length $r$ and of the 
distance $Z$ of the O$_2$ center of mass from the surface. The angles and lateral positions are indicated in the insets. The energy zero (green/yellow border) corresponds to a free triplet O$_2$ molecule. Contour lines are drawn at 0.2~eV intervals. Dissociation barriers (if present) are labeled (eV).}
\end{figure*}

Before discussing the results obtained with this approach, we remind of two 
general problems of present-day Kohn-Sham-DFT: First, even with gradient 
corrected xc functionals the description of the binding energy of the free O$_2$ 
molecule is rather bad. Going from the O$_2$ spin-triplet ground state to two 
free oxygen atoms, each of them also in the spin-triplet ground state, the 
errors of our calculated binding energies with respect to the experimental
value (5.1\,eV \cite{herzberg52}) are: 2.3 eV (LDA), 1.0 eV (PBE), 0.6 eV 
(BLYP), and 0.5 eV (RPBE). Fortunately, for the part of the PES, that is
important for the sticking coefficient, we find that different functionals 
give results that differ by much less, indicating some favorable error 
cancellation. Below we will therefore restrict our discussion to the PBE 
and the RPBE, since they represent the extreme cases for the gradient corrected 
functionals, yielding the strongest and smallest overbinding in the O$_2$ 
molecule, respectively. A second noteworthy problem arises because the 
expectation value of $S^2$ is not defined in Kohn-Sham-DFT. For the present 
case this implies that the multiplet structure is not well 
described~\cite{gunnarsson80,vonbarth79}. In free O$_2$ the many-body ground 
state belongs to the triple degenerate ${^3}\Sigma_{\rm g}^-$ state which 
is followed by two singlets, namely a doubly degenerate ${^1}\Delta_{\rm g}$ 
level (0.98 eV above the ground state), and a non degenerate ${^1}\Sigma_{\rm 
g}^+$ level (1.63 eV above the ground state). While the total energy of the 
spin-triplet ground state is described well, the ${^1}\Delta_{\rm g}$ and 
${^1}\Sigma_{\rm g}^+$ states are not described appropriately, since here DFT 
with jellium-based xc functionals describes a certain mixture of multiplets. 
A reasonable approximation to the true spin-singlet state is instead obtained 
by a spin-unpolarized calculation~\cite{tobepublished}, which is for PBE 
1.1\,eV (for RPBE 1.2\,eV) higher than the spin-triplet ground state.

Figure \ref{pes} shows two cuts through the calculated six-dimensional 
PESs for three situations: the adiabatic approximation (discussed in the
introduction), the spin-triplet PES (using constrained DFT) and the 
spin-unpolarized calculation, which is the best we can do to describe the
spin-singlet PES. Whereas the two elbow plots of the adiabatic PES (cf. Fig. 
\ref{pes} left panels) do not exhibit sizeable energy barriers toward
dissociative adsorption, we find clear barriers on the triplet PES (cf. Fig. \ref{pes} middle panels). In fact, inspecting the whole six-dimensional
triplet PES there is always an energy barrier (the lowest one is 0.05 eV). The 
right panels of Fig. \ref{pes} show the corresponding cuts through the
singlet PES, which never exhibits any energy barriers. Clearly, an O$_2$ 
molecule prepared in the singlet state would therefore react most
efficiently with the Al(111) surface. Since the spin forbidden transition to
the triplet ground state can only proceed by scattering with another molecule,
the long lifetime of a singlet O$_2$ should render molecular beam experiments
possible to verify this proposition.

\begin{figure}
\scalebox{0.3}{\includegraphics{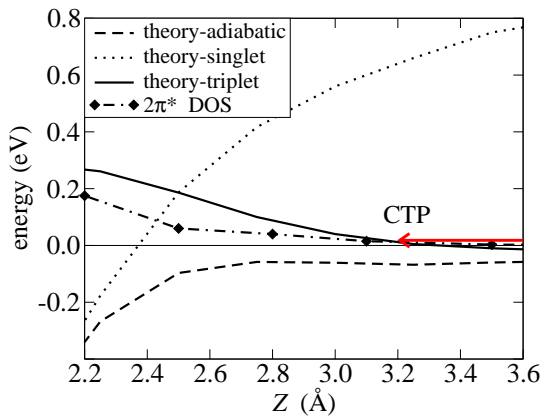}} 
\caption{\label{crossing} Potential energy along the reaction path shown as 
dashed line in Figs. \ref{pes}a, b, c (solid line = triplet PES, dotted line = 
singlet PES, dashed line = adiabatic PES). The red arrow indicates the classical 
trajectory of a thermal O$_2$ molecule constrained to the triplet PES, with CTP 
marking the classical turning point. At this point the coupling, represented by 
the width of the O $2p$ Kohn-Sham level (dash-dotted line), is only just 
emerging.}
\end{figure}

The sticking coefficient for these PESs is calculated as described above, 
i.e., using the ``divide and conquer'' approach \cite{gross97,lorenz04,tobepublished}. The results for the adiabatic and 
the triplet PESs, using the RPBE functional, are given in Fig.
\ref{sticking}. Obviously, the spin-triplet PES gives a sticking curve 
in good agreement with the experimental result. However, when the O$_2$
and Al(111) wave functions overlap at close distances, spin transfer will 
occur with a certain probability. Due to the uncertainty in the
description of the singlet-PES, it is at present not very meaningful to 
perform a quantitative evaluation of these transition probabilities.
A rough estimate of the importance of transitions bringing the system away 
from the triplet-PES is instead provided by the width of the $2\pi^*$ Kohn-Sham 
resonance, which is the level that carries the spin. At large distance the 
width is zero, and it gradually increases upon approach to the surface. 
For a one-dimensional cut through configuration space this is shown in Fig. 
\ref{crossing}. The peak width remains quite narrow and even at the point where 
the triplet and singlet PESs cross it is only about 0.1\,eV. In general, the lifetime of the $2\pi^*$ electrons should be compared to the time the molecule spends between the classical turning point (CTP) and ca. 5 \AA{} away from 
the surface. For thermal molecules (cf. the arrow and the CTP point in Fig. \ref{crossing}) the comparison is: lifetime $\approx$ 3 ps vs. time of 
presence $\approx$ 1 ps. We therefore conclude that for thermal O$_2$ 
molecules (and even for all molecules with a kinetic energy below $\sim 
0.2$\,eV) transitions away from the triplet PES will not play a big role. Our 
results then suggest that particularly these lowest energy molecules should be 
repelled by the barriers on the triplet PES, well before there is significant 
hybridization of wave functions, i.e. before relaxation towards the adiabatic 
ground state occurs. Only for higher kinetic energies, transitions will 
gradually become important, leading to higher sticking coefficients than in
the ``theory-triplet'' curve shown in Fig. \ref{sticking}. We also note 
that the PESs of the PBE and RPBE functionals are similar, but quantitatively 
differences exist. These differences have noticeable influence on the calculated 
sticking curve only for kinetic translational energies below 0.2\,eV. As 
the RPBE gives a better description for free O$_2$ we place a higher credibility 
on its PES. Details will be discussed elsewhere \cite{tobepublished}.

Analyzing the approaching O$_2$ molecule in greater detail reveals finally 
another interesting feature. For molecules that approach in an orientation
perpendicular to the surface (or close to this) the spin is shifted to the atom 
that is further away from the surface. We believe this to be the onset of 
adsorption by the abstraction mechanism. In this way one O atom can adsorb in a 
singlet state, while the spin is efficiently carried away with the other 
O atom that is either repelled back into the vacuum or to a distant place 
at the surface. Calculating the full dynamics of this process, i.e. going
beyond the onset of dissociation important for the sticking coefficient,
requires the explicit consideration of forces on the Al atoms, which we
are implementing at present.

In summary, we have shown that spin selection rules can play an important role 
for O$_2$ scattering at metals. They imply that O$_2$ molecules should travel
in a spin-triplet configuration up to distances close to the surface where 
hybridization with metal-surface states becomes significant. This is
particularly important for systems with a low DOS at the Fermi level; for
transition metals we expect that the high density of $d$-states at the
Fermi level can more easily take up the spin. At Al(111) spin selection 
leads to a very low sticking probability for thermal O$_2$ molecules in
the triplet ground state, while O$_2$ molecules prepared in the singlet 
configuration should adsorb with high probability. Similar effects as those 
discussed in this paper should just as well play a role for other substrates 
with a low jellium-like density of states at the Fermi level, 
and for other molecules with a
high-spin ground state.

\end{document}